\documentclass[%
 aip,
 amsmath,amssymb,
 reprint,%
]{revtex4-1}

\usepackage{graphicx}
\usepackage{dcolumn}
\usepackage{bm}

\usepackage[utf8]{inputenc}
\usepackage[T1]{fontenc}
\usepackage{mathptmx}
\usepackage{etoolbox}
\usepackage{textcomp}

\makeatletter
\def\@email#1#2{%
 \endgroup
 \patchcmd{\titleblock@produce}
  {\frontmatter@RRAPformat}
  {\frontmatter@RRAPformat{\produce@RRAP{*#1\href{mailto:#2}{#2}}}\frontmatter@RRAPformat}
  {}{}
}%
\makeatother
\begin{document}

\preprint{AIP/123-QED}

\title[Anisotropic Exchange Spin Model to Investigate the Curie Temperature Dispersion of Finite-Size L1$_0$-FePt Magnetic Nanoparticles]{Anisotropic Exchange Spin Model to Investigate the Curie Temperature Dispersion of Finite-Size L1$_0$-FePt Magnetic Nanoparticles}

\author{Kohei Ochiai}
\email{ochiai.kohei.xiqrk@resonac.com}
\affiliation{Resonac Corporation, Research Center for Computational Science and Informatics, 8, Ebisu-cho, Kanagawa-ku, Yokohama, Kanagawa. 221-8517, Japan}

\author{Tomoyuki Tsuyama}
\affiliation{Resonac Corporation, Research Center for Computational Science and Informatics, 8, Ebisu-cho, Kanagawa-ku, Yokohama, Kanagawa. 221-8517, Japan}

\author{Sumera Shimizu}
\affiliation{Resonac Corporation, Research Center for Computational Science and Informatics, 8, Ebisu-cho, Kanagawa-ku, Yokohama, Kanagawa. 221-8517, Japan}

\author{Lei Zhang}
\affiliation{Resonac Hard Disk Corporation, Research \& Development Center, 5-1, Yawatakaigan-dori, Ichihara, Chiba, 290-0067, Japan}

\author{Jin Watanabe}
\affiliation{Resonac Hard Disk Corporation, Research \& Development Center, 5-1, Yawatakaigan-dori, Ichihara, Chiba, 290-0067, Japan}

\author{Fumito Kudo}
\affiliation{Resonac Hard Disk Corporation, Research \& Development Center, 5-1, Yawatakaigan-dori, Ichihara, Chiba, 290-0067, Japan}

\author{Jian-Gang Zhu}
\affiliation{Data Storage Systems Center, Carnegie Mellon University, Pittsburgh, PA 15213, USA}
\affiliation{Electrical and Computer Engineering Department, Carnegie Mellon University, Pittsburgh, PA 15213, USA}
\affiliation{Materials Science and Engineering Department, Carnegie Mellon University, Pittsburgh, PA 15213, USA}

\author{Yoshishige Okuno}
\affiliation{Resonac Corporation, Research Center for Computational Science and Informatics, 8, Ebisu-cho, Kanagawa-ku, Yokohama, Kanagawa. 221-8517, Japan}

\date{\today}%

\begin{abstract}
  We developed an anisotropic spin model that accounts for magnetic anisotropy 
  and evaluated the Curie temperature ($\textit{T}_{\mathbf{c}}$) dispersion due to finite size effects in \textit{L}1$_0$-FePt nanoparticles.
  In heat-assisted magnetic recording (HAMR) media, a next-generation magnetic recording technology, 
  high-density recording is achieved by locally heating \textit{L}1$_0$-FePt nanoparticles near their $\textit{T}_{\mathbf{c}}$ and rapidly cooling them. 
  However, variations in $\textit{T}_{\mathbf{c}}$ caused by differences in particle size and shape can compromise recording stability and areal density capacity, 
  making the control of $\textit{T}_{\mathbf{c}}$ dispersion critical.
  In this study, we constructed atomistic LLG models 
  to explicitly incorporate the spin exchange anisotropy of L1$_0$-FePt, based on parameters determined by first-principles calculations.
  Using this model, we evaluated the impact of particle size on $\textit{T}_{\mathbf{c}}$ dispersion.
  As a result, 
  (1) the $\textit{T}_{\mathbf{c}}$ dispersion critical to the performance of HAMR can be reproduced, whereas it was previously underestimated by isotropic models
  and (2) approximately $70\%$ of the experimentally observed $\textit{T}_{\mathbf{c}}$ dispersion can be attributed to particle size effects. 
  This research highlights the role of exchange anisotropy in amplifying finite-size effects and underscores the importance of size control in HAMR media.
\end{abstract}

\maketitle

\section{Introduction}
In recent years, with the rapid expansion of cloud services, big data, and AI, 
the volume of data generated worldwide is predicted to rise dramatically \cite{idc:data-age-2025}. 
Consequently, from the viewpoint of cost and energy efficiency per bit, 
there is a strong demand for increasing the capacity and density of storage systems geared toward data centers.
Conventional magnetic recording techniques face limits in recording density, commonly called the "magnetic trilemma (Trilemma)." 
As a next-generation technology, Heat-Assisted Magnetic Recording (HAMR) has attracted considerable attention \cite{1,2}. 
By locally heating a nanoparticle to near its Curie temperature and rapidly cooling it, 
HAMR enables high-density recording while maintaining large magnetic anisotropy.

A key parameter that influences HAMR performance is the magnetic particles' Curie temperature ($\textit{T}_{\mathbf{c}}$). 
Materials such as \textit{L}1$_0$-phase FePt (hereafter referred to as \textit{L}1$_0$-FePt), with extremely large magnetic anisotropy constants, are especially promising \cite{3,4,5}. 
However, experimental studies report substantial variation in $\textit{T}_{\mathbf{c}}$ due to particle size, shape, and degree of order \cite{9,10,11,12,13,FePt_Mc}. 
This spread in $\textit{T}_{\mathbf{c}}$ can undermine recording stability and degrade the signal-to-noise ratio (SNR) 
by causing mismatch between the writing temperature and each particle's actual Curie temperature \cite{14, SNR}. 
Moreover, there are reports that $\textit{T}_{\mathbf{c}}$ dispersion is crucial for determining areal density capability (ADC) \cite{ADC_distribution}, 
underscoring the need to control it in HAMR media design.
Although $\textit{T}_{\mathbf{c}}$ dispersion's impact on SNR and ADC is understood, which parameters predominantly create such dispersion remains unclear. 
The strong exchange anisotropy of the layered \textit{L}1$_0$-FePt structure \cite{6,7} and real particle shapes have not been fully integrated into theoretical models. 
While some prior studies link particle-size distribution to $\textit{T}_{\mathbf{c}}$ variation \cite{12}, 
it has also been noted that single-domain or isotropic-mesh simulations cannot capture exchange anisotropy effects effectively \cite{8}.

Against this background, the this study aims to incorporate Fe-Fe interactions 
(particularly those arising from exchange anisotropy due to the layered structure) 
obtained from first-principles calculations (DFT) into an atomistic-scale spin model. 
Furthermore, we conduct numerical simulations accounting for realistic particle-size distribution. 
Specifically, we first use a modified simple-cubic (modified-sc) model \cite{12} to assess how $\textit{T}_{\mathbf{c}}$ drops with particle size. 
Next, we build an anisotropic spin model incorporating \textit{L}1$_0$-FePt exchange parameters to reproduce and verify $\textit{T}_{\mathbf{c}}$ dispersion arising from particle-size distribution quantitatively.
This approach captures the $\textit{T}_{\mathbf{c}}$ dispersion more accurately previous isotropic models often underestimate it-and can explain approximately $70\%$ of experimentally reported values. 
We also propose design strategies for mitigating $\textit{T}_{\mathbf{c}}$ dispersion in HAMR media.

\section{Theories}
This section describes the theoretical models and methods employed to evaluate the magnetic properties of \textit{L}1$_0$-FePt nanoparticles quantitatively. 
We first construct an atomistic spin model incorporating the markedly different exchange interactions operating within and between atomic layers (intralayer vs. interlayer). 
We then outline how we extracted the exchange interaction constants using first-principles (DFT) calculations. 
In addition, to facilitate a more straightforward evaluation of the temperature dependence and finite-size effects, we also discuss an approach that combines our atomistic simulations with a lattice-site-resolved mean-field model.

\subsection{Atomistic spin model incorporating anisotropic exchange interactions}
\begin{figure*}
  \includegraphics[width=\linewidth]{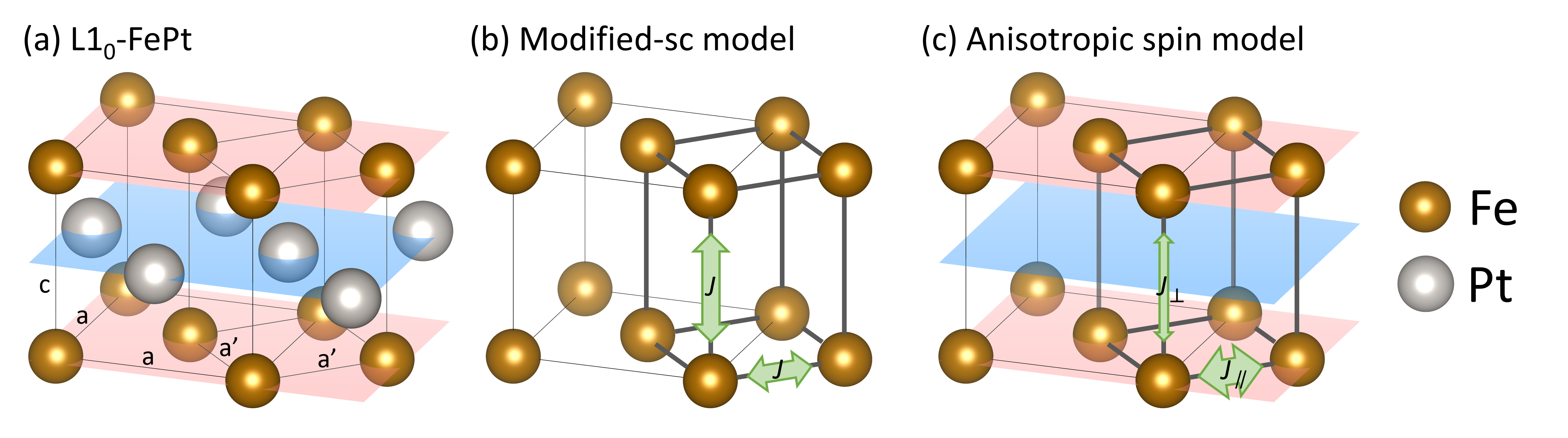} 
  \caption{\label{fig:0} 
  (a) A schematic illustration of the \textit{L}1$_0$-FePt crystal structure. Fe and Pt atomic layers are alternately stackedalong the [001] direction. 
  (b) The modified-sc model proposed by Binh et al. \cite{12}. 
      In the Modified-sc Model, \textit{L}1$_0$-FePt is approximated solely by the Fe degrees of freedom and mapped onto a "simple cubic (sc)-like" lattice with dimensions a'$\times$a'$\times$c. 
      This model simplifies the calculations by assuming isotropic Fe-Fe exchange interactions and unifying the treatment of exchange interactions.  
  (c) The anisotropic spin model proposed in this study 
      considers the in-plane Fe-Fe exchange interaction (\textit{J}$_\parallel$) and the interlayer Fe-Fe exchange interaction (\textit{J}$_\perp$)independently,
      thereby incorporating anisotropic effects arising from the crystal structure.
      }
  \end{figure*}

In \textit{L}1$_0$-FePt, Fe and Pt atomic layers are alternately stacked along the [001] direction \cite{6,7} (Figure \ref{fig:0} -(a)). 
As a result, the exchange coupling within a single Fe layer (in-plane direction, \textit{J}$_\parallel$) is large due to direct Fe-Fe interactions, 
whereas the exchange coupling between adjacent Fe layers (out-of-plane direction, \textit{J}$_\perp$) is weaker because it is mediated via Fe-Pt-Fe pathways \cite{15,16,17,18}. 
Such pronounced exchange anisotropy is critical in the magnetization-reversal mechanism and thermal spin fluctuations.

In previous studies, \textit{L}1$_0$-FePt has often been modeled-primarily for computational efficiency-by a "modified simple cubic" (modified-sc) structure that includes only the Fe degrees of freedom \cite{12,19} (Figure \ref{fig:0} -(b)). 
The DFT studies by Mryasov \textit{et al}. \cite{19} indicated that it is possible to reproduce the overall magnetic properties without explicitly including the Pt layers. 

However, this modeling approach does not explicitly incorporate the substantial exchange anisotropy arising from the layered structure. 
In the design and development of HAMR media, the lateral grain dimension is carefully controlled, further underscoring the significance of anisotropy. 
Consequently, an anisotropic crystal model is essential for accurately evaluating finite-size effects in \textit{L}1$_0$-FePt nanoparticles.

To address this issue, we further refine the modified-sc model by explicitly incorporating the anisotropic exchange interactions (\textit{J}$_\parallel$/\textit{J}$_\perp$) derived from DFT calculations. 
This approach enables us to accurately evaluate finite-size effects and capture the critical role of exchange anisotropy in \textit{L}1$_0$-FePt nanoparticles.

The Hamiltonian for the anisotropic spin model can be written as:
\begin{eqnarray}
H =&& {-} \frac{1}{2} \left[ \sum_{\langle i,j \rangle \parallel} J_\parallel \left( \mathbf{s}_i \cdot \mathbf{s}_j \right) + \sum_{\langle i,j \rangle \perp} J_\perp \left( \mathbf{s}_i \cdot \mathbf{s}_j \right) \right] \nonumber\\
  && {-} k_u \sum_i \left( \mathbf{s}_i \cdot \hat{e} \right)^2 {-} \sum_i \mu_i \left( \mathbf{s}_i \cdot \mathbf{B} \right),
\end{eqnarray}

where $\sum_{\langle i,j \rangle \parallel}$ denotes the summation over nearest-neighbor sites within the same atomic layer, and $\sum_{\langle i,j \rangle \perp}$ denotes the summation over nearest-neighbor sites in adjacent layers. 
The quantity $\mathbf{s}_i$ is the unit spin vector at site $i$, $k_u$ is the uniaxial anisotropy constant, $\mu_i$ is the magnetic moment associated with the spin, and $\mathbf{B}$ is the external magnetic field. 
By increasing the ratio $J_\parallel / J_\perp$, we explicitly account for differences between the in-plane and out-of-plane exchange couplings.

\subsection{Extraction of the ratio of exchange constants via first-principles calculations}
\begin{figure*}
\includegraphics{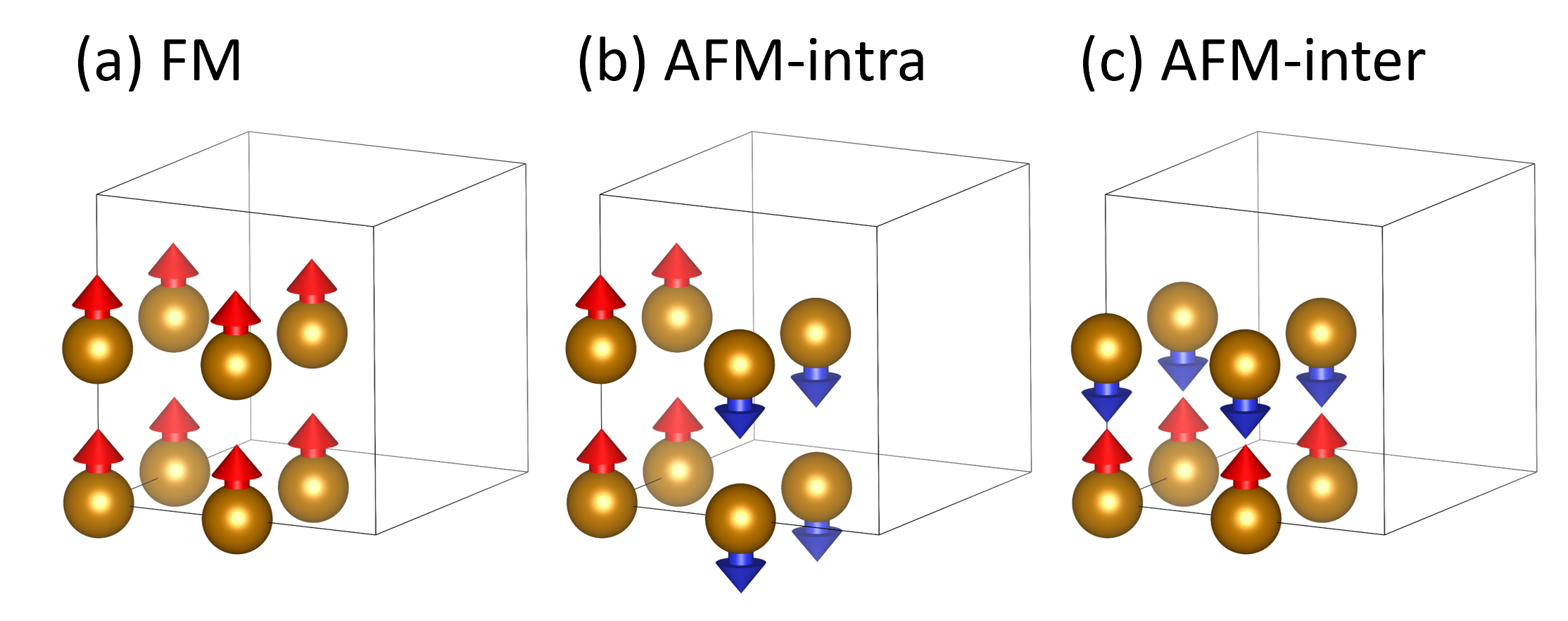}
\caption{\label{fig:3} Examples of the three cell configurations used to estimate the anisotropy of J: (a) FM, (b) AFM-intra, (c) AFM-inter. 
    Fe atoms and their spins are illustrated; Pt atoms are omitted from the schematic. 
    After structural optimization in QE, the total energies of each configuration were calculated.}
\end{figure*}

To quantitatively determine the ratio of intralayer (\textit{J}$_\parallel$) and interlayer (\textit{J}$_\perp$) exchange interactions in \textit{L}1$_0$-FePt, we performed DFT calculations based on the methodology of Xiang et al. \cite{15}. 
Concretely, using Quantum ESPRESSO \cite{36}, we mapped the total energies onto the classical Heisenberg model to extract the coupling constants. 
By switching the spin configurations between ferromagnetic (FM) and antiferromagnetic (AFM) states-either 
within a single atomic layer (AFM-intra) or between layers (AFM-inter) we calculated the following energy differences as schematically illustrated in Figure \ref{fig:3} :

\begin{equation}
\Delta E_\parallel = E(\text{AFM-intra}) - E(\text{FM}),
\end{equation}
\begin{equation}
\Delta E_\perp = E(\text{AFM-inter}) - E(\text{FM}),
\end{equation}

These energy differences ($\Delta \textit{E}$) are used to determine the exchange constant \textit{J} 
by considering the number of spin sites (\textit{N}) and the magnitude of each spin (\textit{S}), 
consistent with the classical Heisenberg model. 
Specifically, by employing the energy difference between FM and AFM, 
one can define the total energy change of the spin system as the spin-spin interaction and thus obtain \textit{J}, as shown in the following equation.

\begin{equation}
  J = \frac{\Delta E}{NS^2},
  \end{equation}

We obtained the respective values of \textit{J}$_\parallel$ and \textit{J}$_\perp$ from these calculations.

\subsection{Temperature-dependent analysis via lattice-site-resolved mean-field modeling}
To complement the atomistic simulations and rapidly estimate the temperature dependence of magnetization and the Curie temperature ($\textit{T}_{\mathbf{c}}$), 
we also employed a mean-field approximation model capable of resolving each lattice site separately \cite{20}. 
In this approach, the effective magnetic field at each site $i$, $\mathbf{B}_\text{eff,i}$, is given by:
\begin{equation}
\mathbf{B}_\text{eff,i} = \sum_j J_{ij} \langle \mathbf{s}_j \rangle + \mu \mathbf{B},
\end{equation}
where $\langle \mathbf{s}_j \rangle$ represents the thermally averaged spin at site $j$. 
Using the Langevin function to evaluate $\langle \mathbf{s}_j \rangle$, we can obtain a qualitative picture of the finite-temperature magnetization dynamics. 
Hence, at a coarse-grained level, this approach enables the estimation of $\textit{T}_{\mathbf{c}}$ and its dependence on finite-size effects, which can be compared directly with simulations based on the atomistic spin model.

\subsection{Calculation of Curie temperature using atomistic spin simulation}

In this study, 
we analyzed the magnetic properties of FePt particles by partially modifying the atomic spin simulation software "VAMPIRE," 
developed by Evans et al. \cite{23,24,25,26}. 
VAMPIRE applies the Metropolis Monte Carlo method to determine the thermal equilibrium distribution of magnetization at temperature $T$ and defines the Curie temperature $\textit{T}_{\mathbf{c}}$ as the point at which the magnetization strength decreases sharply with increasing temperature.

More specifically, as the temperature is varied, the average magnetization length and the average magnetic susceptibility are obtained, 
and the following approximate equation, incorporating the critical exponent $\beta$, is used to determine $\textit{T}_{\mathbf{c}}$:
\begin{equation}
m(T) = \langle \sqrt{\sum_i \mathbf{S}_i} \rangle = (1 - T/\textit{T}_{\mathbf{c}})^\beta,
\end{equation}

Here, $\mathbf{S}_i$ denotes the spin vector on site $i$, and $m(T)$ is the normalized magnetization length. 

Additionally, for the determined $\textit{T}_{\mathbf{c}}$, 
we analyzed its particle size dependence by fitting the model represented by the following equation(Eq. \ref{eq:TcFit}) using the Finite-Size Scaling Law (FSSL) \cite{27}.

\begin{equation}
\label{eq:TcFit} 
\textit{T}_{\mathbf{c}}(D) = T_{c0} \left( 1 - x_{01} D^{-1/\nu_1} \right),
\end{equation}

Where  $D$ is the particle width, T$_{c0}$ is the Curie temperature for the bulk material, and $x_{01}$ and $\nu_1$ are fitting parameters. 
When both the width $D$ and the height $h$ of the particle are varied, this expression is extended accordingly.

\section{Results and discussion}
\subsection{Evaluation of the Curie Temperature's Finite-Size Effect Using the Modified-sc Model}

\subsubsection{Overview and Significance}

In this section, we reintroduce the "modified-simple cubic (modified-sc) model" introduced by Binh $\textit{et al}$. \cite{12} 
and adapt it for our FePt nanoparticles to investigate the finite-size effect on the Curie temperature (Tc). 
Our objectives are (i) to verify whether this model, under our chosen conditions, can replicate the experimental trend of Tc as a function of particle size, 
and (ii) to clarify the model's limitations, which will form the basis for the extended approach presented in Part B.

\subsubsection{Details of the Calculation Method}

First, a "modified-sc" model proposed by Binh $\textit{et al}$. \cite{12} was introduced. 
In this model, the simulation restricts the exchange interaction to only nearest neighbors, assuming it isotropic. 
The exchange interaction parameter $J_{ij}$ is fitted such that the theoretical Curie temperature of bulk FePt (660 K)\cite{28} is reproduced. 
As a result, we obtained $J_{ij} = 6.303 \times 10^{-21}$ J/link. 
In addition, the normalized atomic spin moment $U_s$ for the Fe atom and the uniaxial anisotropy constant $K_u$ are set to 3.23 $\mu_B$ and 2.63 $\times 10^{-22}$ J/atom, respectively \cite{29}.

Although the modified-sc model significantly simplifies the treatment of exchange interactions, it still retains reasonable consistency with experimental and theoretical values. 
This model offers an extremely rational approach, especially for this study, which aims to evaluate the Curie temperature, magnetization distribution, and microscopic behavior across a variety of particle sizes, heights, and shapes in a rapid and qualitative manner.

\subsubsection{Calculation Results}

In this study, simulations were carried out on square-column FePt crystal grains with various particle sizes and heights. 
Using the aforementioned FSSL, the particle size dependence of the Curie temperature was analyzed.

(1) Fixing the particle height at 10 nm and varying only the particle width

\begin{figure}
\includegraphics[width=\columnwidth]{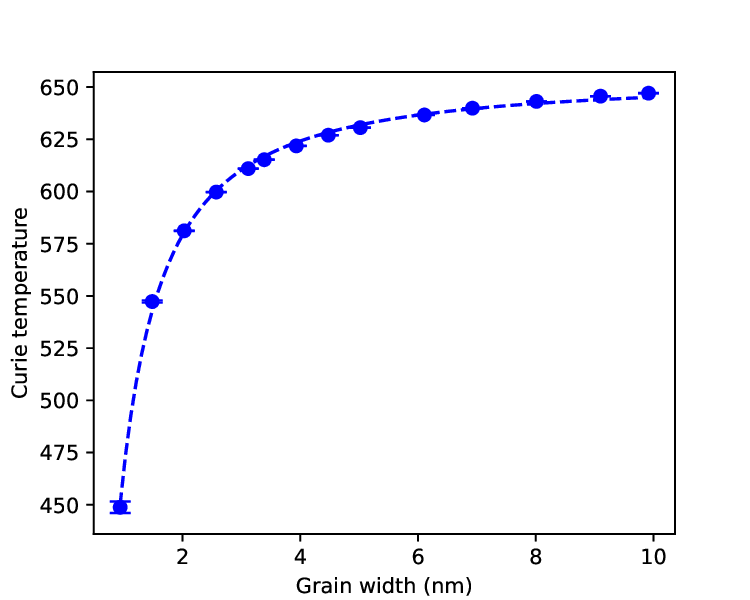}
\caption{\label{fig:1} Curie temperature as a function of particle width (with height fixed at 10 nm), along with the FSSL fitting. Around 5 nm, Tc begins to converge.}
\end{figure}

As shown in Figure \ref{fig:1}, $\textit{T}_{\mathbf{c}}$ increases sharply in the small particle diameter regime and tends to converge once the width is approximately 5 nm or greater. 
By applying FSSL for fitting, good agreement with the simulated data is achieved.

(2) Fixing the particle width at 6 nm or 8 nm and varying only the height

\begin{figure}
\includegraphics[width=\columnwidth]{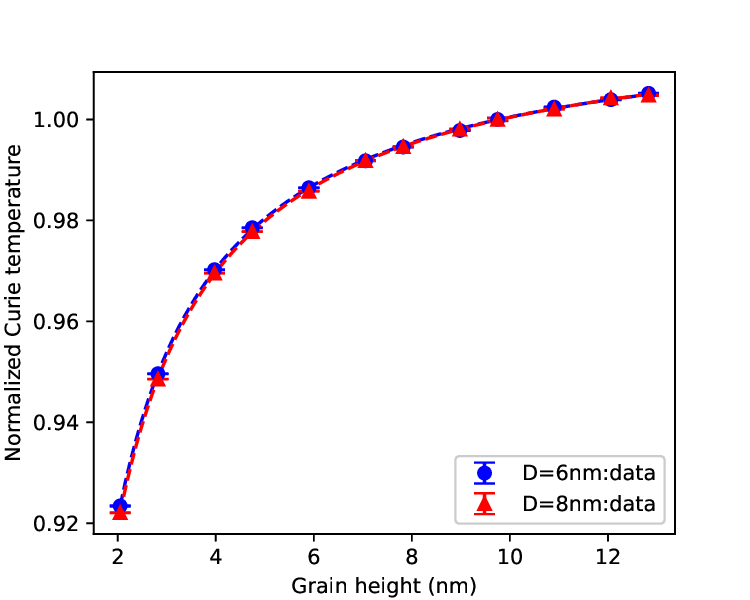}
\caption{\label{fig:2} Curie temperature versus particle height for widths of 6 nm and 8 nm, normalized to the value at h = 10 nm. Both examples exhibit similar behavior and fit well to the FSSL formula.}
\end{figure}

Figure \ref{fig:2} illustrates the behavior of $\textit{T}_{\mathbf{c}}$ when the height is varied. 
The results for 6 nm and 8 nm widths both show similar trends, again in good agreement with the FSSL fitting. 
When normalized by the value of $\textit{T}_{\mathbf{c}}$ at the height of 10 nm, the height-dependent variation becomes more evident.

From Figures \ref{fig:1} and \ref{fig:2}, it is clear that changes in both the width and height of the particles yield size-dependent variations in $\textit{T}_{\mathbf{c}}$ that follow the FSSL. 
Consequently, to incorporate width $D$ and height $h$ simultaneously into the finite-size effect, we perform the following extended fitting:

\begin{equation}
\label{eq:dTc} 
\textit{T}_{\mathbf{c}}(D,h) = T_{c0} \left( 1 - x_{1} D^{-1/\nu_1} \right) \left( 1 - x_{2} h^{-1/\nu_2} \right)
\end{equation}
The seven fitting parameters obtained from this analysis are summarized in Table \ref{tab:1}.
\begin{table*}
\caption{\label{tab:1} Representative fitting parameters for Eq. \ref{eq:dTc}}
\begin{ruledtabular}
\begin{tabular}{ccccccc}
$T_{c0}$ & $x_1$ & $\nu_1$ & $x_2$ & $\nu_2$ \\
\hline
$670.5 \pm 1.4$ & $0.247 \pm 0.002$ & $0.844 \pm 0.018$ & $0.194 \pm 0.001$ & $1.026 \pm 0.013$ \\
\end{tabular}
\end{ruledtabular}
\end{table*}

In recent studies, 
it has been reported that real FePt particles with an average particle diameter of approximately 7-9 nm and $\sim15\%$ particle-size distribution 
exhibit a Curie temperature dispersion $\sigma \textit{T}_{\mathbf{c}}$ of approximatedly $2\%$ \cite{30,31,32,33,34,35}. 
Similarly, in media evaluations within Resonac Corporation, it has been observed that, for instance, the sample in Table \ref{tab:2} has approximatedly a $2.5\%$ spread in $\textit{T}_{\mathbf{c}}$, with a $20\%$ spread in width and a $12\%$ spread in height.

\begin{table*}
  \caption{\label{tab:2} Example of Resonac in-house media evaluation results}
  
  \begin{ruledtabular}
  \begin{tabular}{cccccc}
  Average Width & Width Dispersion & Average Height & Height Dispersion & Curie Temperature & $\sigma \textit{T}_{\mathbf{c}}$  \\
  \hline
  $7.8$ nm  & $20\%$ & $10$ nm & $12\%$ & $650$ K & $2.5\%$ \\
  \end{tabular}
  \end{ruledtabular}
  \end{table*}

Using Eq. \ref{eq:dTc} in the present modified-sc model, 
we estimate the size-dependent $\textit{T}_{\mathbf{c}}$ dispersion $\sigma \textit{T}_{\mathbf{c}, \text{size}}$ for the same particle size and height distribution referenced in Table \ref{tab:2} via:
\begin{equation}
\sigma \textit{T}_{\mathbf{c}, \text{size}} = \frac{\textit{T}_{\mathbf{c}}(\text{max}) - \textit{T}_{\mathbf{c}}(\text{min})}{2 \times \textit{T}_{\mathbf{c}}(\text{central})}  
\end{equation}
where $\textit{T}_{\mathbf{c}}(\text{max})$ and $\textit{T}_{\mathbf{c}}(\text{min})$ are the Curie temperatures for particles whose sizes are (mean ± distribution), and $\textit{T}_{\mathbf{c}}(\text{central})$ is the value for the mean size. 
Under the conditions in Table \ref{tab:2}, the modified-sc model indicates that the average Curie temperature  $\textit{T}_{\mathbf{c}}$ was found to be 645.7 degrees,
and the Curie temperature dispersion $\sigma\textit{T}_{\mathbf{c}, \text{size}}$ attributable to the size effect alone was approximatedly $0.8\%$, which is smaller than the experimentally observed 2.5

\subsubsection{Discussion}
In the first set of simulations, the size-effect contribution to $\sigma\textit{T}_{\mathbf{c}, \text{size}}$ was calculated to be $0.8\%$, considerably lower than the experimental value of 2.5\%. 
If the size effect accounts for only $\sim0.8\%$, then the remaining $\sim1.7\%$ is presumably attributable to other factors, such as variations in ordering (particularly the degree of \textit{L}1$_0$ layering) or lattice strain.

The significance of particle size effects on Curie temperature has been pointed out in various previous works \cite{5}, and even if the estimated $0.7\%$ itself appears small, it is still non-negligible. 
However, in \textit{L}1$_0$-FePt, the remarkable exchange anisotropy arising from the layered stacking of Fe and Pt along the [001] direction means that the in-plane (Fe{-}Fe) and out-of-plane (Fe{-}Pt{-}Fe) exchange strengths differ significantly \cite{6,7,15,16,17,18}. 
Because the modified-sc model treats these differences isotropically, it suggests that the finite-size effect on Tc may be underestimated. 

Given these considerations, the following section proposes and evaluates a new spin model incorporating the anisotropy of intra-layer and inter-layer exchange constants, aiming to improve simulation accuracy.

\subsection{Development of an Anisotropic Spin Model}

\subsubsection{Details of the Calculation Method}

To address the issues identified in the previous section, 
we extended the modified-sc model to develop a new atomic spin model ("anisotropic spin model) that incorporates exchange anisotropy. 
We employed the VAMPIRE software to conduct atomic spin simulations, explicitly distinguishing between intra-layer and inter-layer exchange interactions. 
In this model, 
the intra-layer exchange interaction is denoted as \textit{J}$_\parallel$ 
and the inter-layer exchange interaction as \textit{J}$_\perp$, 
and their ratio, \textit{J}$_\parallel$/\textit{J}$_\perp$, is referred to as the "Anisotropy Ratio," which can be freely adjusted.
In this section, we set the anisotropy values to 1 (isotropic), 2, 4, and 10, and evaluated the variations in the Curie temperature for each setting. 
First, we calculated the $\textit{T}_{\mathbf{c}}$ of the bulk crystal using the same procedure as the modified-sc model, obtaining approximately 685 K under the new parameter set. 
By adjusting the anisotropy ratio, the parameters listed in Table \ref{tab:various_ani} reproduced the bulk $\textit{T}_{\mathbf{c}}$. 

\begin{table}
  \caption{\label{tab:various_ani} Example of the Vampire parameters for each level of anisotropy ratio in the anisotropic spin model}
  \begin{ruledtabular}
  \begin{tabular}{ccc}
  Anisotropy Ratio & $J_\parallel$ & $J_\perp$ \\
  \hline
  $ 1.0 $ & $6.303 \times 10^{-21}$ & $6.303 \times 10^{-21}$ \\
  $ 2.0 $ & $7.733 \times 10^{-21}$ & $3.867 \times 10^{-21}$ \\
  $ 4.0 $ & $9.064 \times 10^{-21}$ & $2.266 \times 10^{-21}$ \\
  $ 10.0 $ & $10.930 \times 10^{-21}$ & $1.093 \times 10^{-21}$ \\
  \end{tabular}
  \end{ruledtabular}
  \end{table}

The spin magnetic moment (3.23 $\mu_B$) and uniaxial anisotropy constant (2.63 $\times 10^{-22}$ J/atom) are identical to those used in the modified-sc model.

\subsubsection{Calculation Results}
\begin{figure}
\includegraphics[width=\columnwidth]{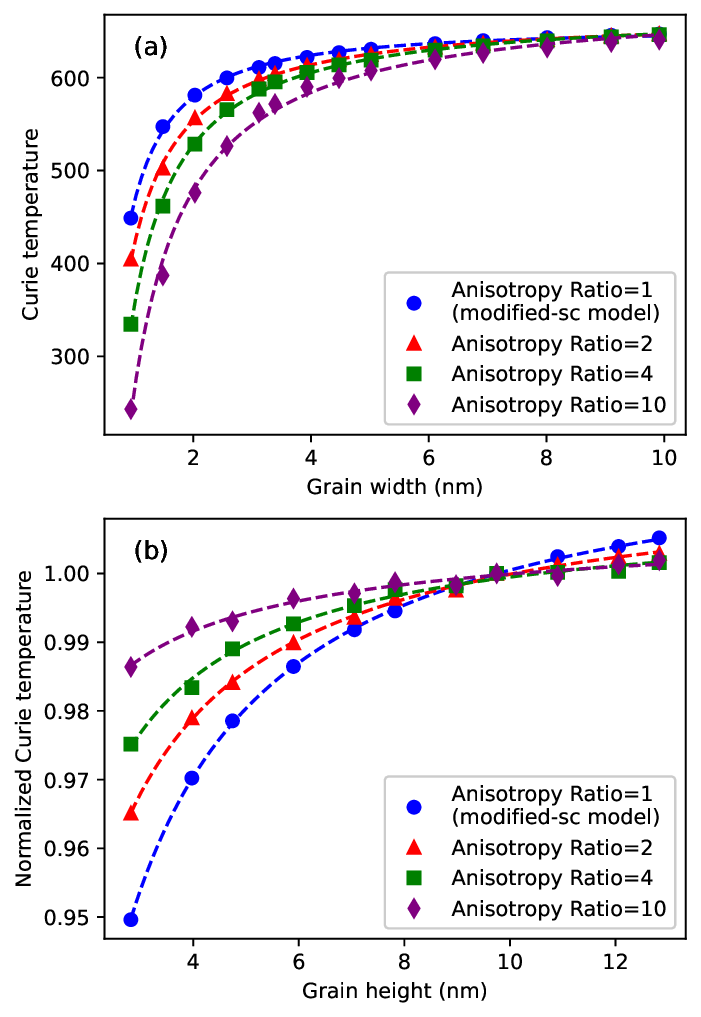}
\caption{\label{fig:various_sigma}  Curie temperature variations for the anisotropic spin model of each anisotropy rartio: 
(a) Particle height is fixed at 10 nm while the width varies. 
(b) Particle width is fixed at 6 nm while the height varies (normalized to the value at h = 10 nm).
}
\end{figure}

Using this new "anisotropic spin model," we calculated the Tc for cubic FePt particles as a function of their width and height, as described in Section A, 
and fitted the results using Eq. \ref{eq:dTc}. 
Figure \ref{fig:various_sigma} shows the $\textit{T}_{\mathbf{c}}$ behavior when (a) the particle height was fixed at 10 nm, 
and the width was varied, and (b) the particle width was fixed at 6 nm, and the height was varied. 
As the degree of anisotropy increases, the Curie temperature variation resulting from changes in particle width becomes significantly more pronounced. 
In contrast, the variation caused by changes in particle height remains relatively small.
In each cases, the Tc behavior followed the Finite Size Scaling Law (FSSL), reaffirming the finite size effect on $\textit{T}_{\mathbf{c}}$.

Furthermore, similar to Section A, we calculated the $\sigma\textit{T}_{\mathbf{c}, \text{size}}$
for each anisotropy level under the size dispersion conditions listed in Table \ref{tab:2}.
The results are shown in Table \ref{tab:various_sigma}
The results indicated that as the anisotropy increased, the impact of width variation on Tc dispersion also increased.

\begin{table}
  \caption{\label{tab:various_sigma} The $\textit{T}_{\mathbf{c}}$ dispersion ($\sigma\textit{T}_{\mathbf{c}, \text{size}}$) calculated for each anisotropy ratio in the anisotropic spin model}
  \begin{ruledtabular}
  \begin{tabular}{ccc}
  Anisotropy Ratio & $\sigma\textit{T}_{\mathbf{c}, \text{size}}$ \\
  \hline
  $ 1.0 $ & $0.8\% $ \\
  $ 2.0 $ & $1.1\%$ \\
  $ 4.0 $ & $1.4\%$ \\
  $ 10.0 $ & $2.0\%$ \\
  \end{tabular}
  \end{ruledtabular}
  \end{table}

\subsubsection{Discussion}
This trend suggests that a larger $\sigma\textit{T}_{\mathbf{c}, \text{size}}$ can be achieved, closely aligning with the Tc dispersion observed in actual experiments. 
The calculations demonstrated that introducing anisotropic exchange interactions leads to more significant variation (dispersion) in Curie temperature due to size effects compared to the modified-sc model. 
When the exchange interactions are treated equivalently both within and between layers, as in the modified-sc model, the $\textit{T}_{\mathbf{c}}$ dispersion tends to be significantly underestimated relative to experimental values. 
However, introducing anisotropy brings the results closer to reality.
From these findings, it is clear that considering the strong exchange anisotropy inherent in the \textit{L}1$_0$-FePt layered structure is essential for accurately estimating the Curie temperature of nanoparticles. 
Therefore, in the next section, we aim to determine the precise ratio of intra-layer to inter-layer exchange constants (anisotropy ratio) for actual FePt crystals through first-principles (DFT) calculations. 
In the subsequent section, we will use these values to quantitatively demonstrate the contribution of size effects to the experimentally observed Tc dispersion.

\subsection{Estimation of the Ratio of Anisotropic Exchange Constants from DFT Simulations}

\subsubsection{Computational Details}
To evaluate the anisotropy ratio of exchange interactions in \textit{L}1$_0$-FePt, we performed DFT calculations using the Quantum ESPRESSO package \cite{36}. 
We extracted the difference between \textit{J}$_\parallel$ and \textit{J}$_\perp$ from these calculations. 
We employed the PBE exchange-correlation functional \cite{37} and Projector Augmented-Wave (PAW) potentials \cite{38} for Fe and Pt. 
The plane-wave cutoff energies were set to 75 Ry for wavefunctions and 600 Ry for charge density, and we used a Monkhorst-Pack k-point mesh of $32 \times 32 \times 32$ points for Brillouin zone integrations. 
The supercell was constructed by alternating Fe and Pt atomic layers along the [001] direction, comprising a total of eight layers (16 atoms). 
Lattice constants and atomic positions were initialized using experimental data \cite{39}, followed by energy minimization.

\subsubsection{Calculation Results}
\label{sec:B2}
To extract the exchange constants, we considered three collinear magnetic configurations and computed the total energy differences $\Delta E$ in each:

(A) FM: All Fe spins aligned along +z.
(B) AFM-intra: Spins alternate between +z and -z within each atomic layer.
(C) AFM-inter: Each layer is ferromagnetic internally, but adjacent layers alternate between +z and -z.

Our calculations yielded:
\begin{equation}
\Delta E_\parallel = 0.003424 \text{ Ry} = 7.466 \times 10^{-21} \text{ J},
\end{equation}
\begin{equation}
\Delta E_\perp = 0.000486 \text{ Ry} = 1.056 \times 10^{-21} \text{ J},
\end{equation}

Given that there are eight Fe spins (S = 1) in the unit cell, the per-link exchange constants \textit{J}$_\parallel$ and \textit{J}$_\perp$ become:

\begin{equation}
J_\parallel = 9.333 \times 10^{-22} \text{ J/link},
\end{equation}
\begin{equation}
J_\perp = 1.320 \times 10^{-22} \text{ J/link},
\end{equation}

indicating that in-plane exchange \textit{J}$_\parallel$ is approximately 7.06 times stronger than out-of-plane exchange \textit{J}$_\perp$.

\subsubsection{Discussion}
These results confirm that the in-plane Fe-Fe exchange is substantially stronger than the out-of-plane Fe-Pt-Fe superexchange. 
Similar conclusions have been reported in earlier work \cite{15,16,17,18}. 
The pronounced exchange anisotropy resulting from the layered structure of \textit{L}1$_0$-FePt strongly affects magnetization reversal and the finite-size behavior of $\textit{T}_{\mathbf{c}}$. 
In the present study, we incorporate this ratio (\textit{J}$_\parallel$/\textit{J}$_\perp$ $\approx$ 7) into the anisotropic spin model 
to quantitatively demonstrate the contribution of size effects to the experimentally observed $\textit{T}_{\mathbf{c}}$ dispersion.

\subsection{Application of the Anisotropic Exchange Interaction Ratio to the Anisotropic Spin Model}
\subsubsection{Parameter Settings}
Based on the anisotropy ratio \textit{J}$_\parallel$/\textit{J}$_\perp$ $\approx$ 7 obtained in Section \ref{sec:B2}, we set the exchange interaction parameters for the anisotropic spin model. 
As before, the $\textit{T}_{\mathbf{c}}$ for the bulk crystal was confirmed to be around 685 K.
After adjusting the ratio \textit{J}$_\parallel : J_\perp = 7 : 1$ , we verified that the parameters in Table \ref{tab:3} reproduced the $\textit{T}_{\mathbf{c}}$ of the bulk system.

\begin{table}
  \caption{\label{tab:3} Example of the Vampire parameters for the anisotropic spin model}
  \begin{ruledtabular}
  \begin{tabular}{ccc}
  Anisotropy Ratio & $J_\parallel$ & $J_\perp$ \\
  \hline
  $ 7.06 $ & $10.033 \times 10^{-21}$ & $1.422 \times 10^{-21}$ \\
  \end{tabular}
  \end{ruledtabular}
  \end{table}

\subsubsection{Curie Temperature Calculation Results}
\begin{figure}
\includegraphics[width=\columnwidth]{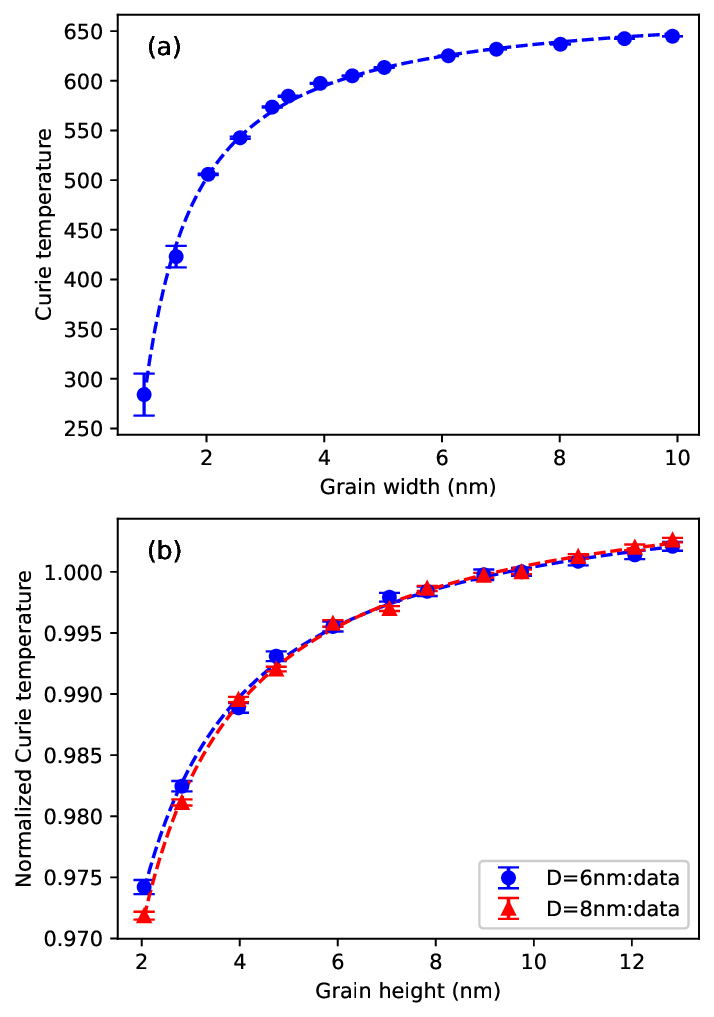}
\caption{\label{fig:4} Curie temperature variations for the anisotropic spin model: 
(a) Particle height is fixed at 10 nm while the width varies. 
(b) Particle width is fixed at 6 or 8 nm while the height varies (normalized to the value at h = 10 nm).
In both cases, good agreement with the FSSL is observed.}
\end{figure}

Using this exchange interaction ratio \textit{J}$_\parallel$/\textit{J}$_\perp$ $\approx$ 7, we replicated the analysis performed in Section A by varying the width and height of cubic FePt grains, 
then fitted the resulting size-dependent $\textit{T}_{\mathbf{c}}$ curves with Eq. \ref{eq:dTc}. 
Figure \ref{fig:4} plots the observed $\textit{T}_{\mathbf{c}}$ variation under the following conditions: 
(a) fixing the particle height to 10 nm while changing the width, and 
(b) fixing widths of 6 nm or 8 nm while varying the height. 
The data follow the FSSL trend in all cases, too.

Representative fitted parameters are summarized in Table \ref{tab:4}. 
Comparing them with their counterparts in the modified-sc model (Table \ref{tab:1}), it is evident that $x_1$ and $x_2$ differ, indicating a change in how strongly $\textit{T}_{\mathbf{c}}$ depends on size.

\begin{table*}
\caption{\label{tab:4} Example of fitting results for the anisotropic spin model}
\begin{ruledtabular}
\begin{tabular}{ccccccc}
$T_{c0}$ & $x_1$ & $\nu_1$  & $x_2$ & $\nu_2$ \\
\hline
$691.5 \pm 6.7$ & $0.488 \pm 0.006$ & $1.045 \pm 0.027$ & $0.067 \pm 0.001$ & $1.107 \pm 0.052$ \\
\end{tabular}
\end{ruledtabular}
\end{table*}

Moreover, for a size distribution like that in Table \ref{tab:2}, the average Curie temperature  $\textit{T}_{\mathbf{c}}$ was found to be 637.8 degrees 
and the size-dependent Curie temperature dispersion $\sigma \textit{T}_{\mathbf{c}, \text{size}}$ attributable to the size effect alone was found to be approximatedly $1.7\%$, 
implying that roughly $70\%$ of the experimentally reported $\sigma \textit{T}_{\mathbf{c}}$ ($2\sim2.5\%$) could be explained by the size effect. 
The remaining $\sim0.8\%$ is presumably due to other contributing factors such as chemical ordering and lattice strain. 
Nevertheless, the improved accuracy of the new model is apparent.

\section{Conclusion}

In this study, we focused on the Curie temperature ($\textit{T}_{\mathbf{c}}$) dispersion of \textit{L}1$_0$-FePt nanoparticles. Specifically, 
(1) we employed a modified simple cubic (modified-sc) model to evaluate the decrease in $\textit{T}_{\mathbf{c}}$ due to finite-size effects, 
and (2) constructed an anisotropic-sc model incorporating \textit{L}1$_0$-FePt exchange parameters to quantitatively reproduce and validate the variation in $\textit{T}_{\mathbf{c}}$ caused by particle size dispersion. 
The principal findings, significance, and remaining challenges are summarized below.

\subsection{Main Findings and Technical Significance}
First, when only the finite-size effect was considered via the modified-sc model, 
the predicted $\textit{T}_{\mathbf{c}}$ dispersion was approximately $0.8\%$, 
falling short of the experimentally reported value of around $2.5\%$. 
By contrast, our anisotropic-sc model, which captures the strong exchange anisotropy inherent in \textit{L}1$_0$-FePt, 
demonstrated that the sizeable in-plane exchange coupling enhanced spin correlation, 
increasing the overall $\textit{T}_{\mathbf{c}}$ dispersion to $1.7\%$, thereby providing better agreement with experimental observations. 
This result demonstrates that including exchange anisotropy dramatically changes the finite-size effect in FePt nanoparticles, 
underscoring the critical importance of realistic modeling for accurately estimating $\textit{T}_{\mathbf{c}}$ dispersion. 

\subsection{Industrial Implications}
This study shows that particle size and size distribution strongly influence $\textit{T}_{\mathbf{c}}$ dispersion. 
Therefore, applications that utilize large assemblies of nanoscale magnetic particles-such as HAMR media-require extremely careful size distribution management. 
On the other hand, 
even the proposed anisotropic-sc model retains a discrepancy of about $0.8\%$ compared to experimental values, 
suggesting that combining additional measures such as enhancing the degree of chemical order and optimizing surface/boundary structures in the manufacturing process could further reduce $\textit{T}_{\mathbf{c}}$ dispersion. 
Such an integrated approach is expected to offer a promising route toward simultaneously achieving thermal stability and high recording density.

\subsection{Future Issues and Prospects}
We confirmed that our newly proposed "anisotropic spin model" is highly promising as a more precise means of predicting $\textit{T}_{\mathbf{c}}$ dispersion and magnetization characteristics in FePt nanoparticles. 
However, further inclusion of real material complexities-such as interactions beyond the nearest neighbor, inhomogeneities within particles, and the addition of other elemental species-will be necessary to achieve closer quantitative agreement with experimental data. 
Moreover, many issues remain, including the need to address local variations in magnetic properties induced by compositional changes near interfaces and quantify incoherent magnetization reversal modes. 
In future works, we plan to continue phase-by-phase verification with first-principles calculations and multi-scale simulations, aiming to develop models with broad applicability and high accuracy.

\par
In summary, the anisotropic spin model that explicitly accounts for the pronounced exchange anisotropy in \textit{L}1$_0$-FePt nanoparticles proves to be highly beneficial for analyzing and designing next-generation HAMR media, 
wherein the layered structure notably amplifies finite-size effects. 
As such, our results are expected to aid both academic research and industrial development, providing guidelines for optimizing FePt-based magnetic materials in heat-assisted magnetic recording and high-efficiency spintronic applications.

\nocite{*}
\bibliography{aip_ochiai_t1}

\end{document}